\documentclass[AMA,STIX1COL]{WileyNJD-v2}
\usepackage{moreverb}

\usepackage{graphicx}
\usepackage{epstopdf}
\usepackage{lipsum}
\usepackage{setspace}

\newcommand{\bea}{\begin{eqnarray}}
\newcommand{\eea}{\end{eqnarray}}
\newcommand{\bes}{\begin{subequations}}
\newcommand{\ees}{\end{subequations}}

\newcommand\BibTeX{{\rmfamily B\kern-.05em \textsc{i\kern-.025em b}\kern-.08em
T\kern-.1667em\lower.7ex\hbox{E}\kern-.125emX}}

\articletype{Regular Article}%

\accepted{\& Published in Physica Scripta: \href{https://doi.org/10.1088/1402-4896/abbca0}{https://doi.org/10.1088/1402-4896/abbca0}}

\begin{document}

\title{Dynamics of Higher-order Bright and Dark Rogue Waves in a New (2+1)-Dimensional Integrable Boussinesq Model}

\author[1]{\bf Sudhir Singh}

\author[2]{\bf Lakhveer Kaur}

\author[3,4]{\bf K. Sakkaravarthi*}

\author[5]{\bf R. Sakthivel}

\author[1]{\bf K. Murugesan}

\authormark{\large Higher-order Rogue Waves in a New (2+1)D Integrable Boussinesq Model \hfill Sudhir Singh {\it et al}\hfill}

\address[1]{\orgdiv{Department of Mathematics}, \orgname{National Institute of Technology Tiruchirappalli}, \orgaddress{\state{Tiruchirappalli -- 620015, Tamil Nadu}, \country{India}}\\}

\address[2]{\orgdiv{Department of Mathematics}, \orgname{\\Jaypee Institute of Information Technology}, \orgaddress{\state{Noida -- 201309, U.P.}, \country{India}}\\}

\address[3]{\orgdiv{Department of Physics}, \orgname{National Institute of Technology Tiruchirappalli}, \orgaddress{\state{\\Tiruchirappalli -- 620015, Tamil Nadu}, \country{India}}\\}

\address[4]{\orgdiv{Centre for Nonlinear Dynamics, School of Physics}, \orgname{Bharathidasan University}, \orgaddress{\state{Tiruchirappalli -- 620024, Tamil Nadu}, \country{India}}\\}

\address[5]{\orgdiv{Department of Applied Mathematics}, \orgname{Bharathiar University}, \orgaddress{\state{\\Coimbatore -- 641046, Tamil Nadu}, \country{India}}}

\corres{*{\bf K. Sakkaravarthi}\\ 
\email{ksakkaravarthi@gmail.com}\\ Tel. +91 98949 82606}


\abstract[Abstract]{\bf \large	This work deals with the dynamics of higher-order rogue waves in a new integrable (2+1)-dimensional Boussinesq equation governing the evolution of high and steep gravity water waves. To achieve this objective, we construct rogue wave solutions by employing Bell polynomial and Hirota's bilinearization method, along with the generalized polynomial function. Through the obtained rogue wave solutions, we explore the impact of various system and solution parameters in their dynamics. Primarily, these parameters determine the characteristics of rogue waves, including the identification of their type, bright or dark type doubly-localized rogue wave structures and spatially localized rational solitons, and manipulation of their amplitude, depth, and width. Reported results will be encouraging to the studies on the rogue waves in higher dimensional systems as well as to experimental investigations on the controlling mechanism of rogue waves in optical systems, atomic condensates, and deep water oceanic waves. }

\keywords{Higher dimensional nonlinear model; Boussinesq Equation; Rogue Waves; Rational Solitons\newline~~\newline {\Large Journal Reference:\newline {\it Physica Scripta} {\bf 95} (2020) 115213.\newline \href{https://doi.org/10.1088/1402-4896/abbca0}{https://doi.org/10.1088/1402-4896/abbca0}}}

\maketitle

\setstretch{1.3}
{\large 
\section{Introduction}\label{sec1}
Oceanic rogue waves are one of the significant nonlinear wave structures arising in the deep sea, and they played an essential role in the naval transport, which has been recorded in various instances \cite{rogue-book}. These rogue waves are nothing but a new kind of nonlinear type coherent structures localized in all dimensions/directions and short-lived in time. Due to this non-trivial behaviour, rogue wave has been defined as `wave coming from nowhere and disappears with no trace' \cite{nak}. Such doubly localized (both in spatial and temporal dimensions) excited wave structures can also be referred to as ``freak waves, extreme waves, monster waves, killer waves, and giant waves" as given in the literature. An important reason for these definitions is their exceptionally high amplitude with multi-fold magnitudes on a steady sea state, which provides the ability to destroy even bigger boats and ships during their transport in the deep sea as well as the oil platforms \cite{mho,spe,ebi,cat,mon}. In contrary to other nonlinear waves that have stable long-living characteristics, for example, solitary waves or solitons, the significance/effect of these rogue waves are much less. Also, due to their less frequent occurrence, they have attracted limited interest until a few decades ago. However, their presence in other physical systems, especially optical communication systems and atomic condensates, ignited interest among the researchers working along with those fields. Since then, numerous results on the dynamics and importance of these rogue waves are being reported regularly, particularly in the past two decades. Further, it has been recreated in an artificial water tank experiment recently by executing the familiar nonlinear Schr\"odinger model \cite{ach,PRX-rogue}. Also, wind-perturbed rogue waves in the hydrodynamics system and an annular wave flume that was also experimented in laboratory \cite{acha,ato}. Additionally, the initially recorded Draupner wave was recently recreated in the laboratory to understand the role of breaking in crossing seas \cite{draupner-2020}. 

Although various observations confirm the existence of rogue waves, still the origin of rogue waves remains a debatable subject \cite{kdy,ckh}. Several theoretical studies and certain experimental investigations showed the modulation instability as a pioneering phenomenon in their generation \cite{aal,rpe}. Further, the synchronization of several coherent structures has also been understood as another aspect of these rogue wave generation. Their behavior is mysterious and it can also be explained with the chaotic phenomenon. Also, it has been shown that rogue waves can appear in a wider range of systems apart from the hydrodynamic models \cite{bki}, including nonlinear optics and lasers \cite{mer}, atmosphere \cite{lst}, plasma physics \cite{hba}, and matter waves (Bose-Einstein condensate) \cite{yvb,jhv}. These investigations have given further impulse, and the interest in rogue waves is now a well-motivated multidisciplinary research area.  Additionally, a steady increase in the occurrence of these monster waves in some ocean regions due to extreme weather because of global warming has led to a focus on their intensive exploration \cite{ebi,cat}. Still, there exist several open questions on their formation and dynamics, that enable several researchers towards a continuous study on these rogue waves for the past two decades \cite{PhysD3,PhysD4,nature,znf17,he21,ohta12}. {It is important to remind certain salient differences between rogue waves and tsunami waves (are one kind of solitary waves). Tsunami waves are series of high-amplitude waves and they washes against the coast several times with great speed and force. Tsunamis occur as a result of the movement of a huge volume of seawater from the seabed (shall be due to the movement of tectonic plates resulting from under water earth quakes) to the sea surface and they are stable for a quite long distance. On the other hand, oceanic rogue waves are caused by the movement of wind over the sea surface and they are highly unstable which results into their rapid disappearance. Apart from the generation mechanism, another important difference between these two is that the rogue waves cause damages in deep sea only over a short distance/duration, whereas the tsunami waves result huge destruction in the coastal region spread over several thousands of kilometers too \cite{rogue-book}.}

Mathematically, the rogue waves can be well defined by a set of nonlinear differential equations of scalar (single) as well as multi-component systems. Such rogue waves also arise in higher-dimensional systems. It is important to note that some of the  (2+1)-dimensional systems have fundamental rogue waves which further contain a line profile known as line rogue waves studied with analytical and numerical methods. There is a considerable difference between the profile of fundamental rogue waves in a (1+1)-dimensional system and that of (2+1)-dimensional systems. The central peak is surrounded by several gradually decreasing peaks in (1+1)-dimensional system which are quite distinct in non-fundamental rogue waves. There are various reports on rogue wave solutions for different evolution equations constructed using different analytical methods, including the famous Inverse spectral transform, Darboux transformation, Hirota method, dressing method, and several ansatz approaches \cite{rogue-book,PhysD3,nature,PhysD4,znf17,he21,ohta12}. 
The results on rogue waves reveal that their dynamics in (2+1)-dimensional systems is a fascinating topic and deserves further investigations in different soliton equations. Due to the crucial role of nonlinearity in ocean wave dynamics, the formation of rogue waves and the difficulty in solving the corresponding nonlinear models by analytical/semi-analytic methods is a tough task. However, the involvement of extensive computation makes it little convenient to take the challenge in obtaining multiple rogue wave solutions. Thus, continuous analysis of the rogue waves will help in the enrichment of a complete understanding of the mysterious phenomenon. The interest of researchers has now been shifted to explore multiple rogue wave solutions in addition to multi-soliton solutions. Interaction between these rogue waves and solitons gives rise to a new kind of solution which is of further interest in recent years \cite{nature}.

One of the fundamental nonlinear wave equations which describes the flow in shallow inviscid layer is the Boussinesq model \cite {jvb}
\begin{equation} \label{eq12}
u_{tt}-u_{xx}-\beta (u^2)_{xx}-\gamma u_{xxxx}=0
\end{equation}
where $\beta$ is the vertical extent of fluid, and $\gamma$ represents the velocity of wave profiles. Several integrable/nonintegrable Boussinesq-type equations are proposed with dispersion, temporality, and nonlinearity to model the various phenomena in coastal areas, oceanic rogue waves and tsunami waves, and analytical solutions are obtained. As our objective in this work is to investigate the dynamics of multiple rogue waves in a higher-dimensional nonlinear model, we consider the following (2+1)-dimensional integrable Boussinesq equations proposed by Wazwaz and Lakhveer governing the gravity waves and collisions of surface water waves \cite {Amw-Lk}:
\begin{equation} \label{eq11}
u_{tt}-u_{xx}-\beta (u^2)_{xx}-\gamma u_{xxxx}+ \frac{\alpha^2}{4} u_{yy}+\alpha u_{yt}=0,
\end{equation}
where $\alpha, \beta$ and $\gamma$ are nonzero constants. The above generalized equation (\ref{eq11}) have two additional terms compared to the classical Boussinesq equation (\ref{eq12}), which correspond to the second order dispersive effect with one spatial term ($u_{yy}$) and another spatio-temporal $(u_{yt})$ dimension. Further, the model  (\ref{eq11}) is found to be integrable in the Painlev\'e sense \cite{Amw-Lk} for arbitrary dispersive coefficient $\alpha^2$. It is important to note that the above equation (\ref{eq11}) shall reduces to different versions of Boussinesq and even Benjamin-Ono type model for different choices of $\alpha$, $\beta$ and $\gamma$. Simply, for an example,  Eq. (\ref{eq11}) takes the classical Boussinesq equation for $\alpha = 0$, $\beta=3$ and $\gamma=1$ studied in \cite{mja} and shall be listed a few more versions too which are discussed in the literature. 
There are certain important results on solitary waves, rational solutions, periodic and lump solutions of a different type of (2+1)-dimensional Boussinesq equation and its simple classical model in recent times \cite{NLD2020,pac,NLD-He,ZNF-He,PhysD4,WazMMA2020,MPLB18,NLD2017,NLD2018,arxiv19,NLD2018a,NLD2018b,spring20} to name a few. However, to the best of our knowledge, only the soliton (solitary wave) solutions are reported for the above integrable Boussinesq model (\ref{eq11}) \cite {Amw-Lk}. Still, solutions and dynamics of various other types of nonlinear coherent structures are not available/reported. So, we devote our analysis to construct rogue wave solutions and a detailed study on their dynamics in this manuscript and leave other nonlinear wave solutions for a future investigation. 

Based on the above, our motive in this work is to bring light on the rogue wave solutions and their dynamical characteristics for the considered equation (\ref{eq11}), since, nowadays, theoretical analysis of such waves has become an integral segment of the field nonlinear sciences. The Hirota bilinear method \cite{rhi,ref18,epjp1,epjp2} is found successful in investigating various nonlinear evolution equations to obtain rogue wave solutions and especially it was adopted for low order rogue wave solutions in these studies. Despite the high difficulty level in exploring multiple rogue waves, still, there are few works of literature on it. The symbolic computation approach and polynomials reported in recent years help to study the multiple rogue wave solutions of such equations. 

This paper is divided into various sections consisting of the development of the bilinear equation by using the Hirota bilinear transformation method in Section \ref{sec2}. Next in section  \ref{sec3}, we construct the one, two and three-rogue wave solutions of Eq.  (\ref{eq11}) by implementing a direct and effective way with an establishment of the generalized  polynomial function for the bilinear equation. Further, discussion on the mechanism of rogue waves by controlling the system and solution parameters are also presented. After giving a few important remarks regarding the present work, we provide a brief conclusion in the final section \ref{sec4}.

\section{Bilinearization of the (2+1)D integrable Boussinesq equation}\label{sec2}
This section is devoted for fabricating bilinear form of the Boussinesq equation (\ref{eq11}) by making use of the Bell polynomials. A crisp description about the Bell polynomials can be studied with the help of reference \cite{ref18} for a better understanding. Using the properties of Bell polynomials and considering the transformation 
\bea  \label{eq13}
u(x,\tau)&=&\frac{3\gamma}{\beta}q_{xx}+u_{0},
\eea 
with $\tau=y+a_{1}t,~ x=x$, and an auxiliary function $q=q(x,\tau)=2 \ln (\mathcal{R}(x,\tau))$, the billinear equation of Eq. (\ref{eq11}) can be obtained as
\begin{eqnarray} \label{eq14}
\left((a_{1}^2+\frac{\alpha^2}{4}+\alpha a_{1}) D_{\tau}^2  -(2\beta u_{0}+1)  D_x^2 -\gamma D_x^4 \right) \mathcal{R} \cdot \mathcal{R} =0.
\end{eqnarray}
Here {$\mathcal{R}(x,\tau)$ denotes a new dependent (transformation) function and }$D$ represents the Hirota bilinear operators defined as \cite{rhi}
\begin{equation}
D_x^aD_{\tau}^b(\Lambda \cdot \Theta)=\left( \frac{\partial}{\partial_x}- \frac{\partial}{\partial_{x'}}\right)^a \left( \frac{\partial}{\partial_{\tau}}- \frac{\partial}{\partial_{{\tau}'}}\right)^b  \Lambda(x,{\tau})\Theta(x',{\tau}')|_{x'=x,{\tau}'={\tau}}.
\end{equation}
{For appropriate forms of the function $\mathcal{R}$, different types of nonlinear wave solutions can be derived and their underlying dynamics in system (\ref{eq11}) can be explored through systematic analyses. Here in this manuscript, we focus only on the construction of rogue wave solution by using the above bilinear form and leave other types of nonlinear wave solutions for a future investigation. }

\section{Rogue Waves: Solution and Dynamics}\label{sec3}
{In this section, we construct rogue wave solution of Boussinesq equation (\ref{eq11}). For this purpose, we are making use of a generalized solution structure for higher-order rogue waves proposed in Ref. \cite{azh} along the lines of \cite{mja,pac} for  Kadomtsev-Petviashvili (KP) type equation (a brief revisit on the generalization process is given in Appendix A), which can be written as}
\bes
\bea 
\mathcal{R} = \mathcal{R}_{r+1} (x,\tau;\lambda, \mu ) &=& R_{r+1} (x,\tau) + 2 \lambda \tau F_r (x,\tau) + 2 \mu x G_r (x,\tau) \nonumber\\ &&+ ( \lambda^2 +\mu ^2 ) R _{r-1} (x,\tau),\label{eq27}
\eea
with
\bea
R_{r} (x,\tau) & = \sum \limits_{n=0} ^{r(r+1)/2} \sum \limits _{i=0}^{n} c_{r(r+1)-2n,2i} x ^{r(r+1)-2n } \tau ^{2i}, \\
F_r (x,\tau) & =  \sum \limits_{n=0} ^{r(r+1)/2} \sum \limits_{i=0}^{n} e_{r(r+1)-2n,2i} x ^{r(r+1)-2n } \tau ^{2i}, \\
G_r(x,\tau) & = \sum \limits_{n=0} ^{r(r+1)/2} \sum \limits_{i=0}^{n} h_{r(r+1)-2n,2i} x ^{r(r+1)-2n } \tau ^{2i},  \label{eq27b}
\eea \label{eq27all}\ees 
where $\lambda, \mu , c_{p,q}, e_{p,q}$ and $h_{p,q}~ (p,q=0,2,4, \ldots , r(r+1))$ are arbitrary real parameters to be obtained. 
{The above form of the transformation function $\mathcal{R}$ results into a generalized solution consists of different powers of $x$, $\tau$ and their combinations that give more number of arbitrary parameters. Here we should mention about the methodologies available for constructing general higher-order localized wave solutions, which include Darboux transformation, Riemann-Hilbert formulation and Gauge Transformation where Lax pair of the given model is necessary. However, to the models whose Lax pairs is not known, construction of an infinite number of solutions becomes a challenging task. To such kind of one- and higher- dimensional nonlinear models the considered/present technique can be successively utilized for constructing various nonlinear wave solutions. Recently, a class of local and nonlocal Alice-Bob models \cite{zzh, wli, zzh2,sr19} are solved by using this approach to obtain the bright/dark rogue waves and rational soliton. 
	This method has not been utilized much for obtaining rogue wave solutions due to its recent proposal and definitely shall attract much attraction hereafter. 
	
	Motivated from the above interesting factors, we are interested to adopt the polynomial function given above in Eq. (\ref{eq27all}) for constructing a generalized higher-order rogue wave solution of the (2+1)D Boussinesq equation (\ref{eq11}) and to discuss their dynamics, which we have provided in the rest of the article.}




\subsection{\textbf{The First Order Rogue Wave}}
{In this subsection, we obtain rogue wave solution of the first order to equation (\ref{eq11}) via the bilinear form  (\ref{eq14}) and the above polynomial function $\mathcal{R}$ for suitably restricting the order parameter $r$ up to a finite order. To construct the first order rogue wave solution, we put $r=0$ in (\ref{eq27all}), which results into the following form of $\mathcal{R}$:}
\begin{equation}
\mathcal{R} = \mathcal{R}_1 (x,\tau) = c_{0,0} + c_{0,2} \tau^2 +c_{2,0}x^2. \label{eq29}
\end{equation}
Without loss of generality, we take $c_{2,0}=1$. Substituting Eq. (\ref{eq29}) into the bilinear form yield the following system of equations arising as the coefficients at different powers of $x$ and $\tau$:
\bes\bea
&& \frac{1}{2}c_{0,2}^2 (2 a_1+\alpha)^2+2c_{0,2}(1+ 2 u _0  \beta)  = 0, \\
&& \frac{1}{2}c_{0,0} c_{0,2} (2 a_1+\alpha)^2-2c_{0,0}(1+ 2 u _0  \beta)-12 \gamma  = 0.
\eea \label{eq210} \ees
Solving the above Eq. (\ref{eq210}), we get the rogue wave parameter as
\begin{equation}
c_{0,2} =  \frac{-4 (1+ 2 u _0  \beta) }{ (2 a_1+\alpha)^2}, \qquad 
c_{0,0}  =  \frac{-3 \gamma }{(1+2 u _0 \beta )} . \label{eq211}
\end{equation}
Therefore, we obtain the explicit solution of bilinear equation (\ref{eq14}) as given below.
\begin{equation}
\mathcal{R}_1=(x-\lambda)^2 - \frac{4 (1+ 2 u _0  \beta) }{ (2 a_1+\alpha)^2} (\tau - \mu ) ^2 - \frac{3 \gamma }{(1+2 u _0 \beta )}. \label{eq212}
\end{equation}
Thus by using the above form of $\mathcal{R}_1$ and bilinear transformation (\ref{eq13}), the first order rogue wave solution of the (2+1)D Boussinesq equation (\ref{eq11}) is obtained as
\begin{equation}
u=u_0 +  \frac{12 \gamma}{\beta } \left (  \frac{ \frac{-3 \gamma }{(1+2 u _0 \beta )}-(x-\lambda)^2- \frac{4 (1+ 2 u _0  \beta) }{ (2 a_1+\alpha)^2} (y+a_1 t - \mu ) ^2 }{{\left({ \frac{-3 \gamma }{(1+2 u _0 \beta )}+(x-\lambda)^2- \frac{4 (1+ 2 u _0  \beta) }{ (2 a_1+\alpha)^2} (y+a_1 t - \mu ) ^2 } \right)^2 }} \right ) .\label{eq213}
\end{equation}

{ The above rogue wave solution describes the dynamics of localized excitations appearing in the considered (2+1)D Boussinesq equation, which is characterized by seven arbitrary parameters $u_0$, $\beta$, $\gamma$, $\lambda$, $\alpha$, $\mu$, and $a_1$.  A categorical analysis of the solution (\ref{eq213}) reveals the fact that these arbitrary parameters contribute to determine the dynamics and manipulation of obtained rogue waves under constraint conditions $2a_1+\alpha\neq 0$ and $1+2u_0 \beta <0$, failing which the solution results into singularity or unbounded structure. The evolution of the constructed solution takes a variety of coherent structures in different dimensional planes ranging from a doubly-localized rogue wave to spatially localized rational solitons (special solitary waves). Particularly, the present solution exhibits a doubly localized rogue wave structure along $x-t$ as well as $x-y$ planes while it admits a rational soliton form in the $y-t$ plane. {These rogue waves can be classified into two types based on the peak intensity/amplitude, namely bright rogue wave and dark rogue wave, where the former exhibit a high-amplitude peak above (and two side-band tails below) the continuous background while the latter appears with a deep-amplitude-hole below (along with two side-band bumps above) the constant background. Such bright and dark type rogue waves are depicted in Figs. 1 and 2. Interestingly, the rational soliton arising in the $y-t$ plane admit various profile/wave structure at different spatial coordinate $x$ starting from single-well and double-well dark rational solitons to single-peak rational soliton. They are demonstrated in Fig. 3 for an easy understanding.} 
	
	\begin{figure}[h] 
		\begin{center} 
			\includegraphics[width=0.8435\linewidth]{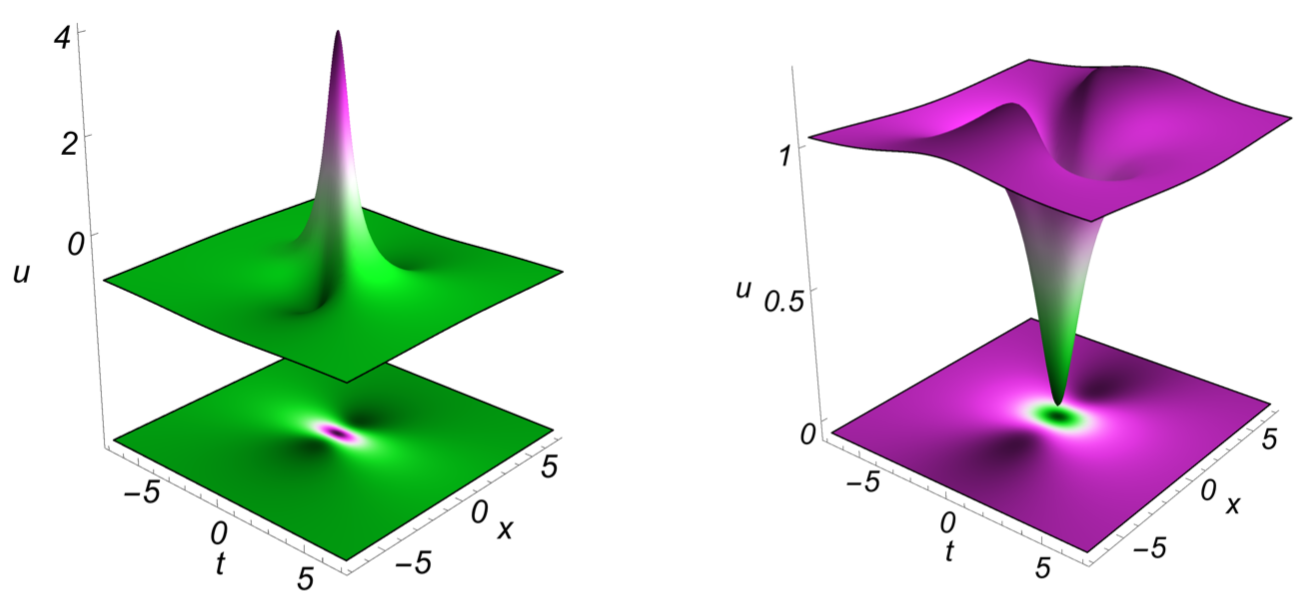}
			\caption{Bright and dark type first order rogue waves through solution (\ref{eq213}). The parameter choice for bright rogue wave is $u_0 = -0.95, \gamma = 0.5, \beta = 1.5,  \lambda = 0.2, a_1 = 1.5, \mu = 0.05$, and $\alpha = 0.5$, while that of the dark rogue wave is $u_0 = 1.04, \gamma = 0.15, \beta = -0.55, \lambda = 0.2, a_1 = -0.35, \mu = 0.5$, and $\alpha = 0.5$ at $y=0.05$.}
		\end{center} 
	\end{figure}
	
	\begin{figure}[h] 
		\begin{center} 
			\includegraphics[width=1.05\linewidth]{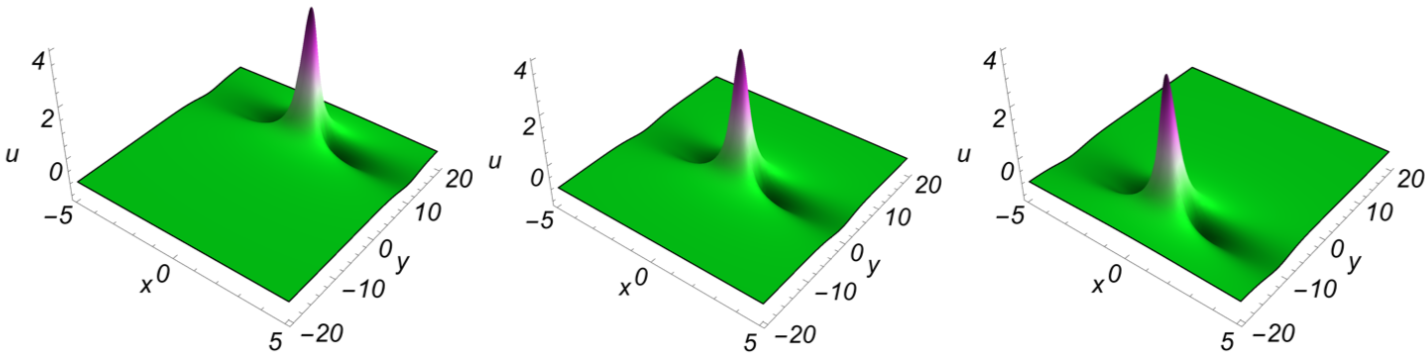}
			\caption{Propagation of bright rogue wave observed at different time $t=-7.5$, $t=0.0$, and $t=7.5$ for $u_0 = -0.95, \gamma = 0.5, \beta = 1.5,  \lambda = 0.2, a_1 = 1.5, \mu = 0.05$, and $\alpha = 0.5$.}  \label{fig-rogue1-2}
		\end{center} \label{fig-rogue1-2}
	\end{figure}
	\begin{figure}[h] 
		\begin{center} \includegraphics[width=0.33\linewidth]{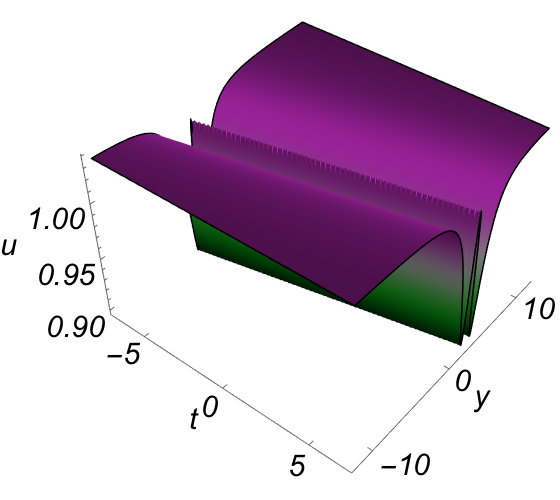}\includegraphics[width=0.33\linewidth]{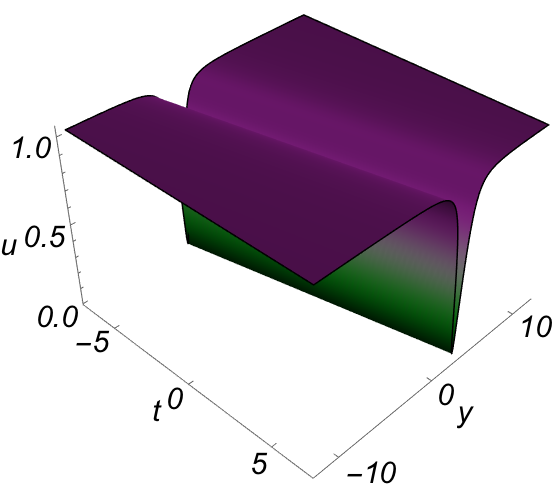}\includegraphics[width=0.36\linewidth]{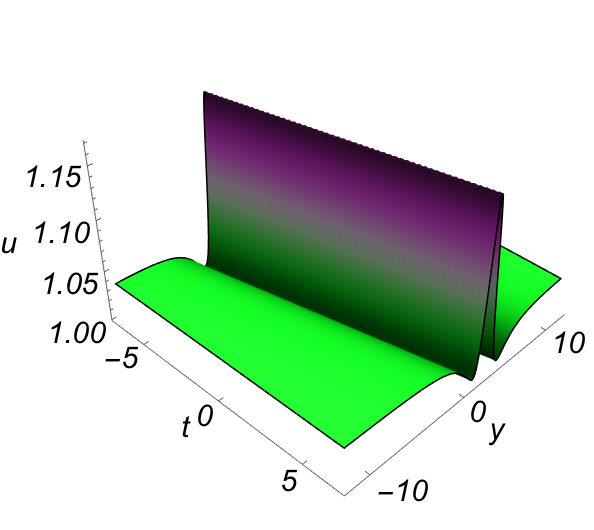}
			\caption{The evolution of rational solitons with various profile structures at different position $x$ for $u_0 = 1.04, \gamma = 0.15, \beta = -0.55, \lambda = 0.2, a_1 = -0.35, \mu = 0.5$, and $\alpha = 0.5$. A W-shaped/double-well and single-well dark solitons and standard rational soliton localized in $y$ at $x=-1.5$, $x=0.05$ and $x=3.5$, respectively. }
		\end{center} \label{fig-rogue1-3}
	\end{figure}
	
	{As mentioned above, the significance of the solution  (\ref{eq213}) is the availability of sufficiently large number of arbitrary parameters and it is necessary to understand their explicit roles in governing the wave dynamics. 
		Importantly, the structure of the rogue wave profile can be controlled by tuning these arbitrary parameters appropriately.} Their impact starts from defining the type/nature of localization, such as bright or dark type rogue waves, to altering other entities like peak-amplitude, depth, and width. Particularly, $u_0$, $\beta$, and $\gamma$ plays a crucial role in determining the nature of wave profile whether the localized excitations becomes bright (intensity peak or amplitude hump) or dark (intensity dip or hole) rogue wave. Additionally, they impact the amplitude peak/depth and width of the rogue waves. Also, $u_0$ provides the background amplitude above/below which the bright/dark rogue wave appears, and it should be a non-zero real constant. The influence of both $a_1$ and $\alpha$ is similar, which are directly proportional to the amplitude as well as the tail-depth. {Another striking feature of $a_1$ parameter is to impart localization along $t$ axis resulting into the formation of rogue wave and rational soliton. Especially, one shall obtain a space-time (doubly) localized bright as well as dark type rogue waves from the solution (\ref{eq213}) when $a_1\neq 0$. Interestingly, the resulting wave becomes a stationary rational soliton (localized in $x$ and $y$ only, but not localized along $t$) when $a_1=0$ in all the three planes $x-t$, $y-t$, and $x-y$.} However, a change in the parameter $\mu$ is inversely proportional/affecting the amplitude and tail-depth. However, it is easy to identify the role of $\lambda$ compared to other parameters, which shifts the position along the $x$ axis. {The above arguments can be visualized/confirmed through Fig. 4, which shall provide a clear understanding on the effects of each arbitrary parameters in bright rogue wave. Similar impact can also be identified in dark rogue wave as well to which the graphical demonstration is not provided here considering the length of the manuscript.} }\\

\begin{figure}[h] 
	\begin{center} 
		\includegraphics[width=0.984\linewidth]{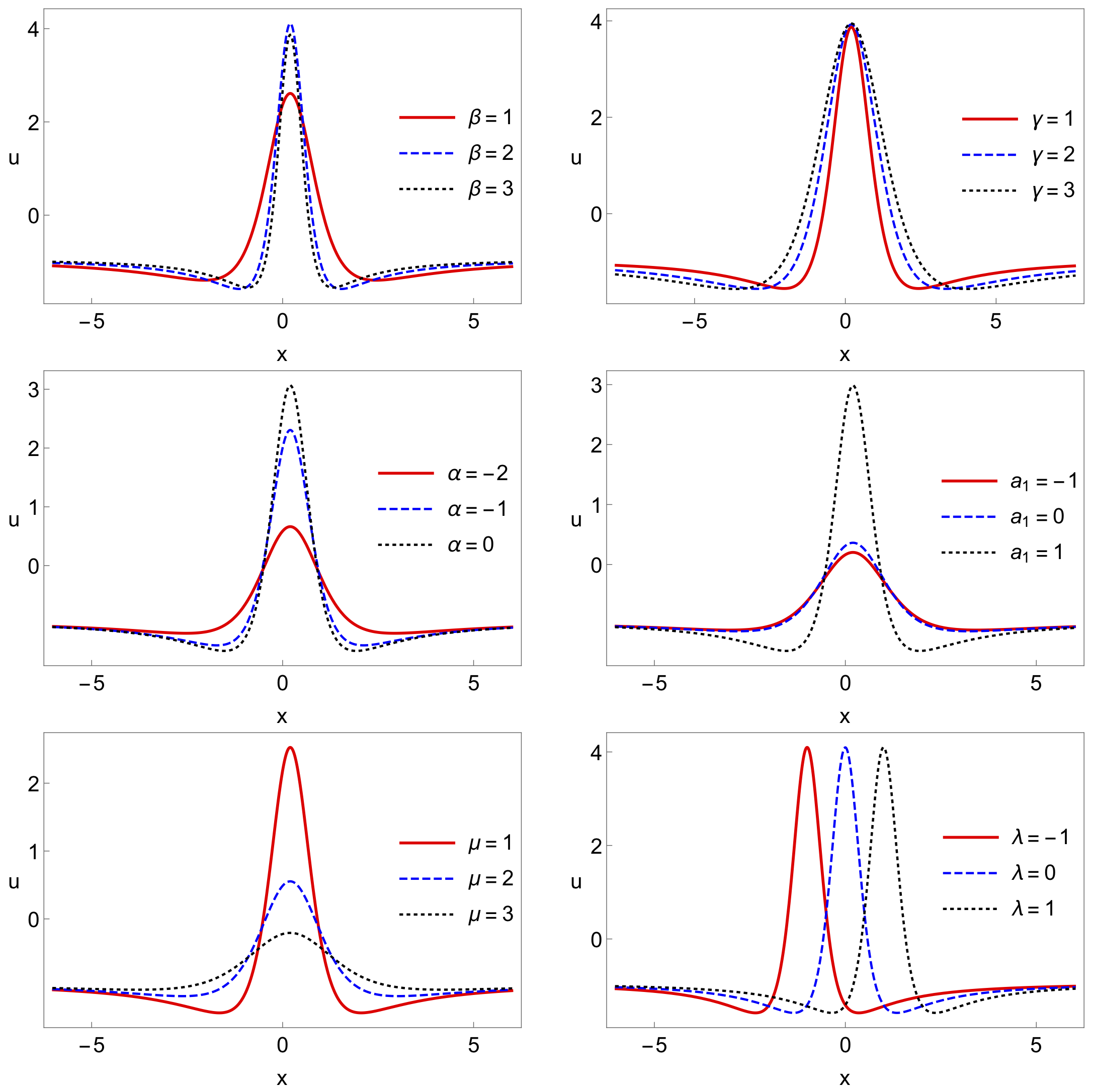}
		\caption{Impact of various arbitrary parameters $a_1$, $\beta$, $\gamma$, $\mu$, $\alpha$, and $\lambda$ in the first order rogue wave (\ref{eq213}) at  $y=0.05$ and $t=0.05$ with other parameters $u_0 = -0.95, \gamma = 0.5, \beta = 1.5,  \lambda = 0.2, a_1=1.5, \alpha = -1.0,$ and $\mu = 0.5$.}
	\end{center} \label{fig-rogue1-4}
\end{figure}
\subsection{\textbf{The Second Order Rogue Wave}}
In continuation to the first order rogue wave solution, here we consider the order parameter $r=1$ in (\ref{eq27all}) to construct the second order rogue wave solution of (2+1)D Boussinesq equation (\ref{eq11}). Thus, we get the approximate form of $\mathcal{R}$ as 
\bes\bea
\mathcal{R} & =& \mathcal{R}_2 (x,\tau;\lambda, \mu )\Rightarrow 
R_2 (x,\tau) + 2 \lambda \tau F_1 (x,\tau) + 2 \mu x G_1 (x,\tau) + ( \lambda^2 + \mu ^2 ) R_0, \nonumber  \\
\mathcal{R}_2&=& ( c_{0,0}+ c_{0,2}\tau^2 + c_{0,4}\tau^4 + c_{0,6}\tau^6 ) + ( c_{2,0} + c_{2,2}\tau^2 + c_{2,4}\tau^4 )x^2  \nonumber  \\ && +(c_{4,0}+ c_{4,2}\tau^2 ) x^4 + x^6 + 2 \lambda \tau (e_{0,0}+e_{0,2}\tau^2 + e_{2,0}x^2 ) \nonumber   \\&& + 2 \mu x (h_{0,0}+ h_{0,2}\tau^2 + h_{2,0}x^2)+ (\lambda^2 + \mu ^2 ) . \label{eq214}
\eea 
Substituting the above $\mathcal{R}$ form into the bilinear equation (\ref{eq14}) and solving all the resultant equations arising at different powers of $x$ and $\tau$, we obtain the following relations among the parameters:
\bea 
c_{0,0}  &=&    \frac{-1}{144(1+ 2 u_0 \beta)^3  } ( 270000 \gamma^3 + (  1+ 2 u _0 \beta ) ( 4 ( 1+ 2 u _0 \beta ) \notag \\ && ( 36 + 4a_1^2 e_{2,0}^2+ 4 a_1 e_{2,0}^2 \alpha + e_{2,0}^2 \alpha ^2 + 72 u _0 \beta ) \lambda^2 -( 16 a_1^4 h_{0,2}^2 + 32 a_1^3 h_{0,2}^2 \alpha  \notag \\
&& + 24a_1 ^2 h_{0,2}^2 \alpha^2 + 8 a_1 h_{0,2}^2 \alpha^3 + h_{0,2}^2 \alpha^4 - 144(1+2 u_0 \beta )^2 ) \mu^2 )) ,   \\
c_{0,2}  &=&   \frac{-1900  \gamma ^2 }{(2a_1+ \alpha )^2 (1+2 u_0 \beta )}, \quad c_{0,4} =  \frac{-272( \gamma+2 u_0 \beta  \gamma )}{(2 a_1 + \alpha )^4 } , \\  
c_{0,6}  & =& \frac{-64(1+2 u_0 \beta )^3 }{(2a_1 + \alpha ) ^6} , \quad c_{2,0}  =\frac{-125 \gamma ^2}{(1+2 u_0 \beta ) ^2 }, \quad  c_{2,2}  =  \frac{360 \gamma  }{(2 a_1 + \alpha )^2} ,\\  
c_{2,4}   &= &  \frac{48(1+2u_0\beta )^2}{(2a_1 + \alpha)^4} , \quad c_{4,0}  =   \frac{-25 \gamma }{1+2 u_0 \beta } ,\quad c_{4,2}   =   \frac{-12 (1+2 u_0 \beta) }{(2 a_1 + \alpha )^2} ,\\ 
e_{0,0} & =& \frac{-5e_{2,0} \gamma}{3(1+2 u_0 \beta )}, \quad e_{0,2} =  \frac{4 (e_{2,0}+ 2 e_{2,0}u_0 \beta )}{3 ( 2 a_1 + \alpha )^{^2} },\\
h_{0,0}  &=& \frac{h_{0,2}(2a_1+\alpha)^2 \gamma}{12(1+2 u_0 \beta ) ^2 }, \quad 
h_{2,0} = \frac{h_{0,2}(2a_1+\alpha)^2}{12(1+2u_0 \beta)}.
\label{eq41}\eea \label{eq41all}
\ees 
From the above parameters explicit form of $\mathcal{R}_2$ required in the bilinear equation (\ref{eq14}) can be obtained. Thus the resultant second order rogue wave solution of (2+1)D Boussinesq equation  (\ref{eq11}) can be constructed straightforwardly from the bilinear transformation  (\ref{eq13}) $u = u_0 +  \frac{6 \gamma }{\beta } ( ln \, \mathcal{ R}_2 ) _{xx}$ with the help of explicit $\mathcal{R}_2$ given above in Eq.  (\ref{eq41all}).
\begin{figure}[h] 
	\begin{center} 
		\includegraphics[width=0.8845\linewidth]{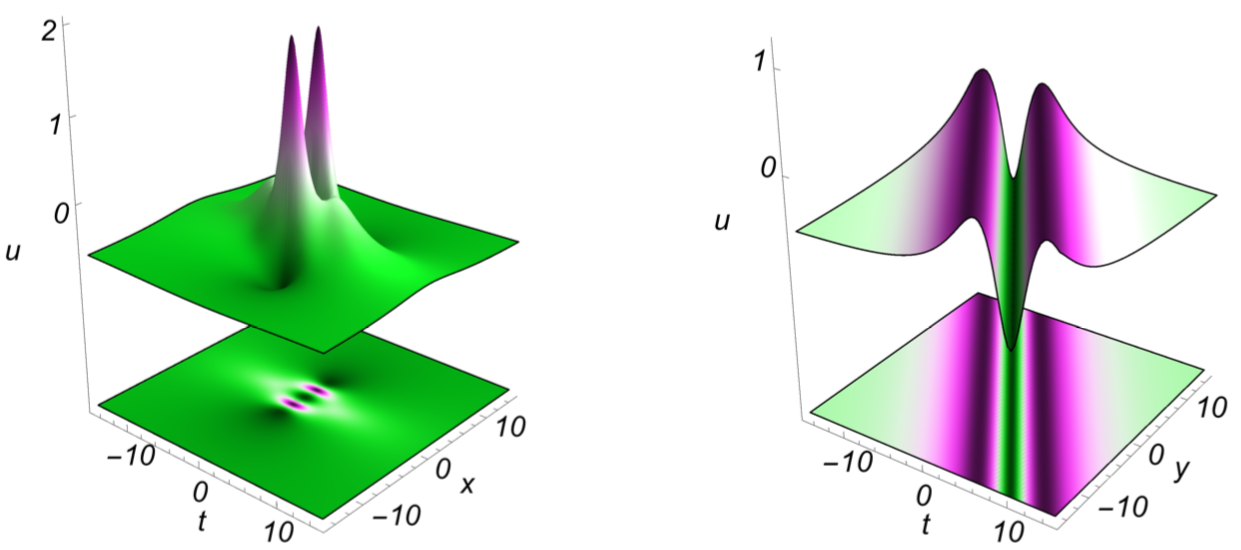}
		\caption{The second order bright rogue wave in x-t plane at $y=0.05$ and M-shaped rational soliton profile in y-t plane at $x=0.05$ obtained through solution (\ref{eq41all}) for $u_0 = -0.59, \gamma= 0.5, \beta = 1.50, \lambda = 0.20, a1 =1.5, \mu = 0.05, \alpha = 0.50, h_{0,2} = 0.2$, and $e_{2,0} = 0.4$.}
	\end{center} \label{fig-rogue1}
\end{figure}

{ The second order rogue wave solution (\ref{eq41all}) consists of nine arbitrary parameters  $u_0$, $\beta$, $\gamma$, $\lambda$, $\alpha$, $\mu$, $a_1$, $h_{0,2}$, and $e_{2,0}$. Here also, one can obtain both bright and dark type rogue waves for appropriate choice of parameters, which consists of doubly localized dual-peak/dip and triple-dip/peak bright/dark profiles, as shown in Figs. 5 and 6. Note that their appearance in $y-t$ and $x-y$ planes takes different structures, wherein the former it admits M-shaped (W-shaped) rational soliton. At the same time, it exhibits a doubly-localized rogue wave profile but with a longer tail in the later $x-y$ plane, respectively, for the bright (dark) solution. Additionally, the explicit impact of each arbitrary parameters can be unveiled, and our analysis shows that they play similar roles as in the case of the first order rogue waves. It starts from the manipulation and control of rogue wave amplitude, width, tail-depth, and localization. {We have shown the impact of only three arbitrary parameters ($u_0,~\beta$, and $\gamma$) in the dark type second order rogue in Fig. 7 and we refrain from the demonstration for other parameters considering the length of the article. It shows that the increase in $u_0$ increases the depth and narrowing (decrease in the width) the dark rogue wave with different background, whereas the opposite scenario occurs for the influence of $\beta$ parameter with unchanged background. However, the parameter $\gamma$ does not alters the amplitude, but it controls/increases the width. Further, the above solution (\ref{eq41all}) also reveals the characteristics of rational solitons (solitary waves) of either propagating one along $y-t$ plane only for $a_1\neq 0$ in addition to the stationary rational structures along all the planes $x-t$, $y-t$, and $x-y$ for $a_1=0$.} For completeness, such stationary rational solitons are also demonstrated in Fig. 8. }
\begin{figure}[h] 
	\begin{center} 
		\includegraphics[width=0.8145\linewidth]{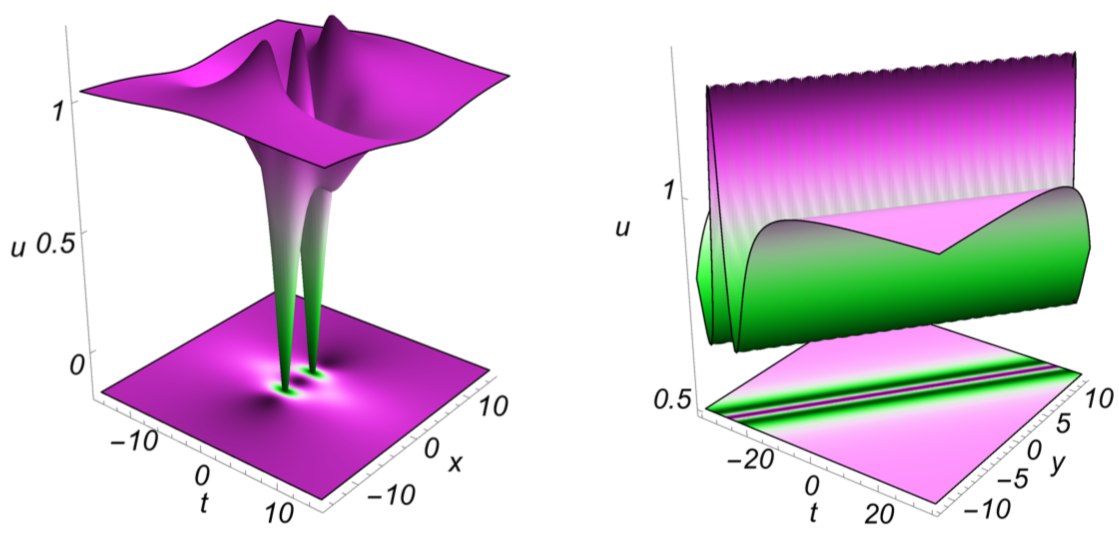}
		\caption{The dark type second order rogue waves in x-t plane at $y=0.05$ and W-shaped rational soliton profile in y-t plane at $x=0.05$ obtained through solution (\ref{eq41all}) for $u_0 = 1.04, \gamma= 0.15, \beta = 0.55,  \lambda = 0.2, a_1 =-0.45, \mu = 0.5, \alpha = 0.5, h_{0,2} = 0.2$, and $e_{2,0} = 0.4$.}
	\end{center} \label{fig-rogue1-3}
\end{figure}
\begin{figure}[h] 
	\begin{center} 
		\includegraphics[width=1.033\linewidth]{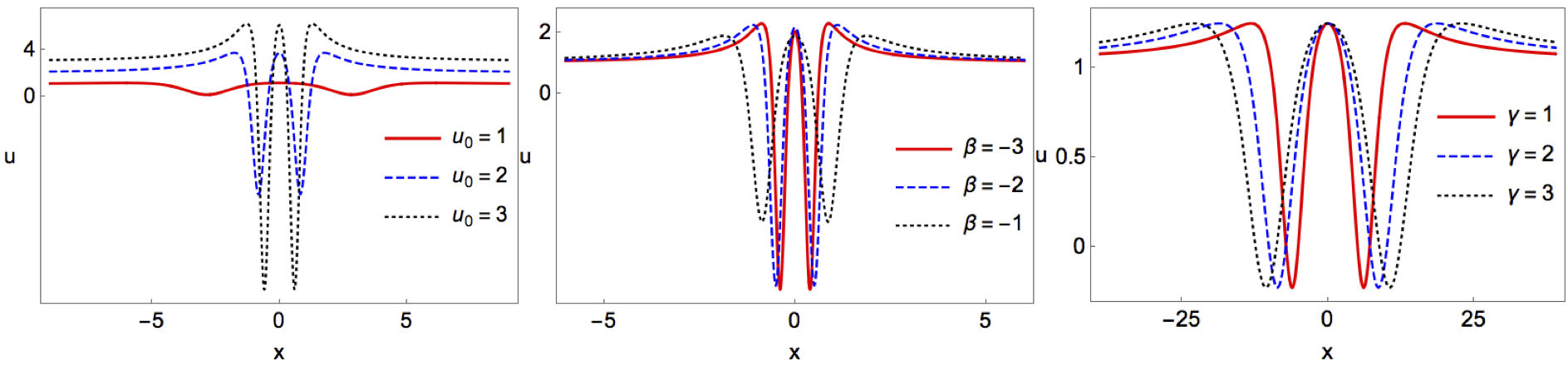}
		\caption{Impact of the arbitrary parameters $u_0$, $\beta$, and $\gamma$ in the dark second order rogue wave (\ref{eq41all}) at  $y=0.05$ and $t=0.05$ with other parameters $u_0 = 1.04; \gamma= 0.15; \beta = 0.55;  \lambda = 0.2; a_1 =-0.45; \mu = 0.5; \alpha = 0.5; h_{0,2} = 0.2$; and $e_{2,0} = 0.4$.}
	\end{center} \label{fig-rogue1-4}
\end{figure}
\begin{figure}[h] 
	\begin{center} 
		\includegraphics[width=0.745\linewidth]{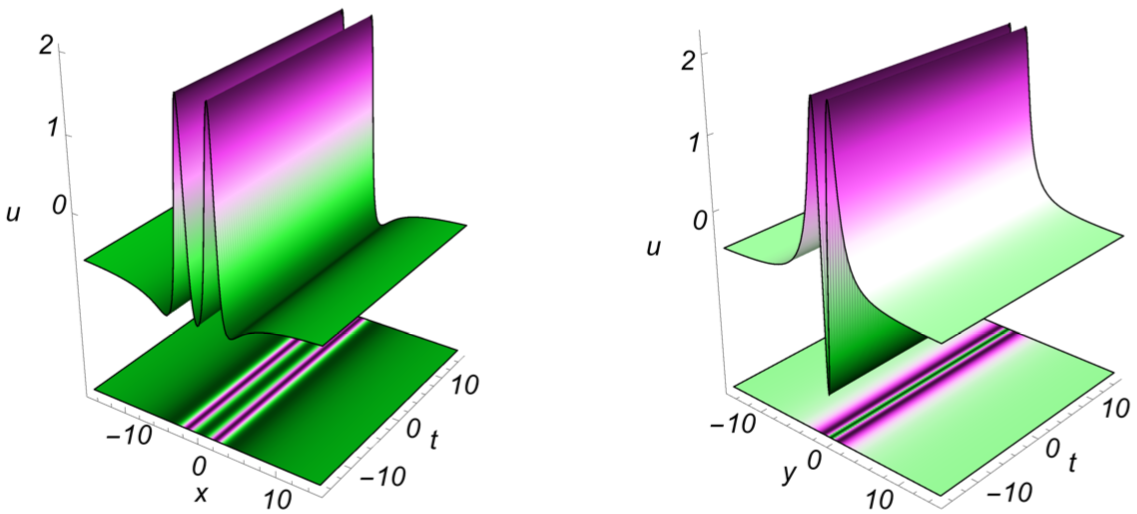}
		\caption{The second order bright rogue wave transforming as a stationary rational soliton (solitary wave) in the x-t and y-t planes with the same parameters as in Fig. 5 except for $a_1=0$.}
	\end{center} \label{fig-rogue1}
\end{figure}
\subsection{\textbf{The Third Order Rogue Wave}}
Extending the algorithm carried out in the first and second order rogue waves, here let us take $r=2$ in (\ref{eq27all}) to extract the third order rouge wave solution of Boussinesq equation (\ref{eq11}), which leads to the form of $\mathcal{R}$ as

\bea \label{eq533}
\mathcal{R} & =& \mathcal{R}_3 (x,\tau;\lambda, \mu )\Rightarrow R_3 (x,\tau) +  2 \lambda \tau F_2 (x,\tau) + 2 \mu x G_2 (x,\tau) + ( \lambda ^2 + \mu ^2 ) R_1 ,  \nonumber\\
\mathcal{R}_3 &=& (c_{0,0} + c_{0,2} \tau ^2 + c_{0,4}\tau ^4 + c_{0,6} \tau ^6 + c_{0,8}\tau ^8 +c_{0,10}\tau ^{10}+c_{0,12}\tau ^{12} \nonumber \\
&&  + (c_{2,0}+ c_{2,2}\tau ^2 + c_{2,4}\tau ^4 + c_{2,6}\tau ^6 + c_{2,8}\tau ^8 + c_{2,10}\tau ^{10}) x ^2 \nonumber  \\ 
&&  + ( c_{4,0} + c_{4,2}\tau ^2 + c_{4,4}\tau ^4 + c_{4,6}\tau ^6 + c_{4,8} \tau^8 ) x ^4 \notag \\
&& + ( c_{6,0}+ c_{6,2}\tau ^2 + c_{6,4}\tau ^4 + c_{6,6}\tau ^6 )x ^6 + ( c_{8,0} + c_{8,2}\tau ^2 + c_{8,4}\tau ^4  ) x ^8 \nonumber \\
&& + ( c_{10,0} + c_{10,2}\tau ^2 ) x ^{10} + x ^{12}  + 2 \lambda \tau (e_{0,0}  + e_{0,2}x^2 + e _{0,4}x^4 + e_{0,6}x^6 \nonumber  \\ 
&&  +(e_{2,0}+e_{2,2}x ^2 + e_{2,4}x ^4 )\tau ^2+  (e_{4,0}+ e_{4,2}x^2)\tau ^4 + \tau ^6 ) \nonumber  \\ 
&& + 2 \mu x \left( h_{0,0}+ h_{0,2}\tau ^2 + h_{0,4}\tau ^4 + h_{0,6}\tau ^6 + (h_{2,0}+h_{2,2}\tau ^2 + h_{2,4}\tau ^4 \right) x ^2 \nonumber  \\ 
&&    + ( h_{4,0} + h_{4,2}\tau ^2 )x^4 + x ^6 ) +  R_1 (x, \tau)  (\lambda^2 +  \mu ^2  ).  \label{rogue3-r3}
\eea 

In a similar manner, by substituting the above $\mathcal{R}$ in the bilinear equation and solving the resulting set of equations at different powers of $x$ and $\tau$, we get the following system of relations among the parameters: 
\bes 
\bea
c_{0,0} & =&   \frac{1}{
	147456 (2 a_1 + \alpha)^ 2 (1 + 2 u_0 \beta )^8} \Big(1769472 a_1 ^{16} \gamma  \lambda ^ 2  + 14155776 a_1 ^{15} \alpha  \gamma  \lambda ^2 + 53 084 160 a_1 ^{14} \alpha ^2 \gamma  \lambda ^2 \nonumber\\
&& \quad  + 123863040 a_1 ^{13} \alpha ^3 \gamma  \lambda ^2 +201277440 a_1 ^{12} \alpha^4 \gamma  \lambda ^2 + 241532928 a_1 ^{11} \alpha ^5 \gamma  \lambda ^2 + 221405184 a_1 ^{10} \alpha ^6 \gamma  \lambda ^2\nonumber\\
&& \quad   + 158146560 a_1 ^9 \alpha ^7 \gamma  \lambda ^2 + 88957440 a_1 ^8 \alpha ^8 \gamma  \lambda ^2 + 39536640 a_1 ^7 \alpha ^9 \gamma  \lambda ^2 + 13837824 a_1 ^6 \alpha ^10 \gamma  \lambda ^2 \nonumber\\
&&\quad  + 3773952 a_1 ^5 \alpha ^{11} \gamma  \lambda^ 2 + 786240 a_1 ^4 \alpha ^{12} \gamma  \lambda ^2 + 120960 a_1 ^3 \alpha ^{13} \gamma  \lambda ^2 + 27 \alpha ^{16} \gamma  \lambda ^2\nonumber\\
&&\quad + 589824 (1 + 2 u_0 \beta )^9 \mu ^2 (\lambda ^2 + \mu ^2)  - 16384 (\alpha  + 2 u_0 \alpha  \beta )^2 (-878826025 \gamma^ 6 - 27 (1 + 2 u_0 \beta )^5 \gamma  \lambda ^2\nonumber \\
&&\quad+ 9 (1 + 2 u_ 0 \beta )^6 \lambda ^2 (\lambda ^2 + \mu ^2))  + 32 a_1  \alpha  (1799835699200 (1 + 2 u_0 \beta )^2 \gamma ^6 + 27 (\alpha^{14} \nonumber\\
&&\quad+ 2048 (1 + 2 u_0 \beta )^7) \gamma  \lambda^ 2 - 18432 (1 + 2 u_0 \beta )^8 \lambda^ 2 (\lambda^ 2 + \mu ^2))  + 32 a_1^ 2 (1799835699200 (1 + 2 u_0 \beta )^2 \gamma ^ 6\nonumber \\
&&\quad+ 27 (15 \alpha ^{14 }+ 2048 (1 + 2 u_0 \beta )^7) \gamma  \lambda  ^2 - 18432 (1 + 2 u_0 \beta )^8 \lambda ^2 (\lambda ^2 + \mu ^2))\Big), \label{rogue3-c00}\\
c_{0,2}& =&\frac{1}{
	12288(2a_1+\alpha)^2(1+2u_0  \beta )^6}
\Big(4929892352000 (1 + 2 u_0  \beta )^2 \gamma ^5 +
3 (16384 a_1 ^{14} + 114688 a_1^{13} \alpha \nonumber \\
&&\quad+ 372736 a_1^{12} \alpha^ 2 + 745472 a_1^{11} \alpha ^3 + 1025024 a_1^{10} \alpha ^ 4 + 1025024 a_1^9 \alpha ^5 + 768768 a_1^8 \alpha ^6 + 439296 a_1^7 \alpha^ 7\nonumber\\
&& \quad  + 192192 a_1^6 \alpha^ 8 + 64064 a_1^5 \alpha ^9 + 16016 a_1^4 \alpha ^{10} + 2912 a_1^3 \alpha ^{11} + 364 a_1^2 \alpha ^{12} + 28 a_1 \alpha ^{13} + \alpha ^{14} \nonumber\\
&& \quad + 16384 (1 + 2 u_0  \beta )^7) \lambda ^2\Big),
\eea  

\bea
c_{2,0} & = &  \frac{1}{49152 (1 + 2 u_0 \beta )^7} \Big( -2617942835200 \gamma ^5 - 10 471 771 340 800 u_0 \beta  \gamma ^5 - 10 471 771 340 800 u_0 ^2 \beta ^ 2 \gamma ^5\nonumber\\
&& \quad  - 49152 \lambda  ^2  - 49 152 a_1^{14} \lambda ^ 2 - 344 064 a_1^{13} \alpha  \lambda^ 2 - 1 118 208 a_1^{12} \alpha ^2 \lambda ^ 2 - 2 236 416 a_1^{11} \alpha ^3 \lambda ^2\nonumber \\
&& \quad - 3 075 072 a_1^{10} \alpha ^4 \lambda^ 2 - 3 075 072 a_1^9 \alpha ^ 5 \lambda ^ 2  - 2 306 304 a_1^8 \alpha ^6 \lambda ^2 - 1 317 888 a_1^7 \alpha^ 7 \lambda  ^2 - 576 576 a_1^6 \alpha ^8 \lambda ^2 \nonumber\\
&&\quad- 192 192 a_1^5 \alpha ^9 \lambda ^2 - 48 048 a_1^4 \alpha ^{10} \lambda ^2 - 8736 a_1^3 \alpha ^{11} \lambda^ 2   - 1092 a_1^ 2 \alpha ^{12} \lambda^ 2 - 84 a_1 \alpha ^{13} \lambda^ 2 - 3 \alpha ^{14 } \lambda^ 2\nonumber\\
&&\quad - 688 128 u_0 \beta  \lambda ^2 - 4 128 768 u_0^2 \beta ^ 2 \lambda ^2 - 13 762 560 u_0^3 \beta ^ 3 \lambda ^2  - 27 525 120 u_0^4 \beta ^ 4 \lambda ^ 2 - 33 030 144 u_0^5 \beta ^5 \lambda ^2 \nonumber\\
&&\quad- 22 020 096 u_0^6 \beta ^6 \lambda ^2 - 6 291 456 u_0^7 \beta ^7 \lambda^ 2\Big) ,\\ 
c_{0,4} &  =&  \frac{262267600 \gamma^ 4}{
	3 (2 a_1 +  \alpha )^4 (1 + 2 u_0 \beta)^2}, \quad 
c_{0,6}   = \frac{51134720 \gamma ^ 3}{
	3 (2 a_1 + \alpha)^6}, \quad 
c_{0,8}  =   \frac{1109760 (1 + 2 u_0  \beta )^2 \gamma ^ 2}{(2 a_1 + \alpha )^8 } , \\ 
c_{0,10} &= &  \frac{59392 (1 + 2 u_0  \beta )^4 \gamma }{(2 a_1 + \alpha )^10} , \quad 
c_{0,12}   = \frac{4096 (1 + 2 u_0 \beta )^6}{(2 a_1 + \alpha )^{12}}  , \quad 
c_{2,2}   =   -\frac{2 263 800 \gamma ^4 } {(2 a_1 + \alpha )^2 (1 + 2 u_0 \beta )^3}  , \\ 
c_{2,4} &  =&  \frac{235 200 \gamma ^ 3}{
	(2 a_1 + \alpha )^4 (1 + 2 u_0 \beta ) } , \quad 
c_{2,6}  =   \frac{-
	2 266 880 (\gamma ^ 2 + 2 u_0 \beta  \gamma  ^2) }{(2 a_1 +  \alpha
	) ^6 }  , \quad 
c_{2,8}   =   \frac{-
	145 920 (1 + 2 u_0 \beta ) ^3 \gamma }{
	(2 a_1 + \alpha )^8}  ,\eea\bea 
c_{2,10} &  =&  \frac{ - 6144 (1 + 2 u_0 \beta  )^5}{(2 a_1 + \alpha )^{10}}  , \quad 
c_{4,0}  =  \frac{- 
	5 187 875 \gamma ^ 4 }{
	3 (1 + 2 u_0 \beta )^4}  , \quad 
c_{4,2}   =   \frac{882 000 \gamma ^ 3}{(2 a_1 + \alpha )^2 (1 + 2 u_0 \beta  )^2}  ,\\
c_{4,4} &  = &  \frac{599 200 \gamma^ 2}{(2 a_1 + \alpha )^4} , \quad 
c_{4,6 } =    \frac{93 440 (1 + 2 u_0 \beta  )^2  \gamma }{(2 a_1 + \alpha )^6 }  , \quad 
c_{4,8}   =   \frac{3840 (1 + 2 u_0 \beta  )^4 }{(2 a_1 +\alpha )^8 }  ,\\
c_{6,0} & = &  \frac{ - 75 460 \gamma ^ 3}{3 (1 + 2 u_0\beta   )^3} , \quad 
c_{6,2 }  =   \frac{- 74 480 \gamma ^ 2 }{(2 a_1 +  \alpha )^2 (1 + 2 u_0 \beta ) } , \quad 
c_{6,4}   =   \frac{-24 640 ( \gamma + 2 u_0  \beta  \gamma )}{(2 a_1 + \alpha )^4 } ,\\
c_{6,6} & =&    \frac{- 1280 (1 + 2 u_0 \beta  )^3}{(2 a_1 + \alpha )^6 }  , \quad 
c_{8,0}   =   \frac{735 \gamma ^ 2 }{(1 + 2 u_0 \beta )^2}  , \quad 
c_{8,2}  =   \frac{2760 \gamma }{
	(2 a_1 + \alpha )^2}  ,\\
c_{8,4} &  =&   \frac{240 (1 + 2 u_0 \beta ) ^2 }{(2 a_1 + \alpha ) ^4 }  , \quad 
c_{10,0} =  \frac{ - 98 \gamma }{1 + 2 u_0 \beta  }   , \quad 
c_{10,2 }   =    \frac{-24 (1 + 2 u_0 \beta )}{(2 a_1 + \alpha )^2}  ,\\ 
e_{0,0} &  = &
\frac{1 }{192 (2 a_1 +  \alpha  )^2 (1 + 2 u_0 \beta )^6 \lambda  }
\Big(4 829 440 a_1^8 \gamma^3 \lambda  + 19 317 760 a_1^7  \alpha   \gamma^3 \lambda  + 33 806 080 a_1^6  \alpha ^ 2 \gamma^3 \lambda\nonumber \\
&&\quad + 33 806 080 a_1^5  \alpha  ^3 \gamma^3 \lambda   +
21 128 800 a_1^4  \alpha ^  4 \gamma^3 \lambda  + 8 451 520 a_1^3  \alpha ^ 5 \gamma^3 \lambda  + 2 112 880 a_1^2  \alpha ^ 6 \gamma^3 \lambda \nonumber \\
&&\quad + 301 840 a_1  \alpha  ^7 \gamma^3 \lambda  + 18 865  \alpha ^ 8 \gamma^3 \lambda  -
768 (1 + 2 u_0 \beta )^7 \mu (\lambda^2 + \mu ^ 2) \Big),\\
e_{0,2}&& =   \frac{665 (2 a_1 + \alpha )6 \gamma ^ 2}{
	64 (1 + 2 u_0 \beta )5 }  , \quad 
e_{0,4} =   \frac{105 (2 a_1 + \alpha )^6 \gamma }{ 
	64 (1 + 2 u_0 \beta ) ^4}
, \quad 
e_{0,6} =   \frac{ -	5 (2 a_1 + \alpha )^6}{64 (1 + 2 u_0 \beta )^3}
,\\
e_{2,0}& =&   \frac{-
	245 (2 a_1 + \alpha )^4 \gamma ^2}{
	16 (1 + 2 u_0 \beta )^4}
, \quad 
e_{2,2} =    \frac{95 (2 a_1 + \alpha )^4 \gamma}{ 
	8 (1 + 2 u_0 \beta )^3}
, \quad 
e_{2,4}=   \frac{ -
	5 (2 a_1 + \alpha )^4}{
	16 (1 + 2 u_0 \beta )^2}
, \\
e_{4,0}& =&   \frac{-
	7 (2 a_1 + \alpha )^2 \gamma  }{
	4 (1 + 2 u_0 \beta )^2}
, \quad 
e_{4,2} =   \frac{9 (2 a_1 + \alpha )^2}{
	4 (1 + 2 u_0 \beta ) },\eea    \bea
h_{0,0 } & =& \frac{
	3 (\lambda  + 2 u_0 \beta  \lambda )^3 - 12 005 \gamma ^3 \mu  + 3 (1 + 2 u_0 \beta )^3 \lambda  \mu^ 2}{
	3 (1 + 2 u_0 \beta )^3 \mu 
} , ~ 
h_{0,2 }  =   \frac{ -
	2140 \gamma ^2}{
	(2 a_1 + \alpha )^2 (1 + 2 u_0 \beta )
}   , ~~\\ 
h_{0,4 } &= &  \frac{ -
	720 (\gamma  + 2 u_0 \beta  \gamma )}{
	(2 a_1 + \alpha )^4
}   , \quad 
h_{0,6 }  =   \frac{ -
	320 (1 + 2 u_0 \beta )^3}{
	(2 a_1 + \alpha )^6
}  , \quad 
h_{2,0 }  =   \frac{ -
	245 \gamma^ 2}{
	(1 + 2 u_0 \beta )^2
}   , \\ 
h_{2,2 } & = &  \frac{ -
	920 \gamma }{
	(2 a_1 + \alpha )^2
}   , ~~ 
h_{2,4 }  =   \frac{ -
	80 (1 + 2 u_0 \beta )^2}{
	(2 a_1 + \alpha )^4
}  ,~~ 
h_{4,0 }  =   \frac{ -
	13 \gamma }{
	1 + 2 u_0 \beta 
}  , ~~
h_{4,2 }  =   \frac{
	36 (1 + 2 u_0 \beta )}{
	(2 a_1 + \alpha )^2
}  . \label{rogue3-h42}
\eea \label{rogue3-par}\ees

Thus the third order rouge wave solution of Boussinesq equation (\ref{eq11}) is obtained by deducing the explicit form of $\mathcal{R}_3$ from the above set of relations  (\ref{rogue3-par}) and substituting it in the bilinear transformation  (\ref{eq13}) $u = u_0 +  \frac{6 \gamma }{\beta } ( ln \, \mathcal{ R}_3 ) _{xx}$.

Our analysis on the above third order rogue wave solution of the considered  (2+1)D Boussinesq equation (\ref{eq11} shows that it contains a number of arbitrary parameters $u_0$, $\beta$, $\gamma$, $\lambda$, $\alpha$, $\mu$, and $a_1$. The role of these parameters in defining the characteristics of the rogue waves is usual, as in the first- and second order of rogue wave dynamics, which includes the alteration of multi-peak-amplitude and multi-hole-depth of the bright and dark rogue waves respectively. In the present third order rogue wave, we have obtained two symmetric peaks on either side of a central maximum peak in the bright type localized wave profile. In contrast, the dark wave structure consists of the merely same number of symmetrically spaced amplitude-holes/dips (lowest amplitude) instead of peaks. This behavior is quite similar in $x-t$ and $x-y$ planes except for the localization in different directions. But, the solution supports a multi-peak (with a central maximum) rational solitons of both bright and dark types along the $y-t$ plane. For illustrative purposes, we have given such bright and dark rogue waves as well as rational solitons in Figs. 9-10. Additionally, one can control $a_1$ parameter to obtain stationary rational solitons as well apart from the moving/traveling rational solitons. {Similar to the first- and second-order rogue wave dynamics, the parameters can be utilized for controlling and manipulating the behaviour of third order rogue waves and rational solitons.} \\ 
\begin{figure}[h] 
	\begin{center} 
		\includegraphics[width=1.05\linewidth]{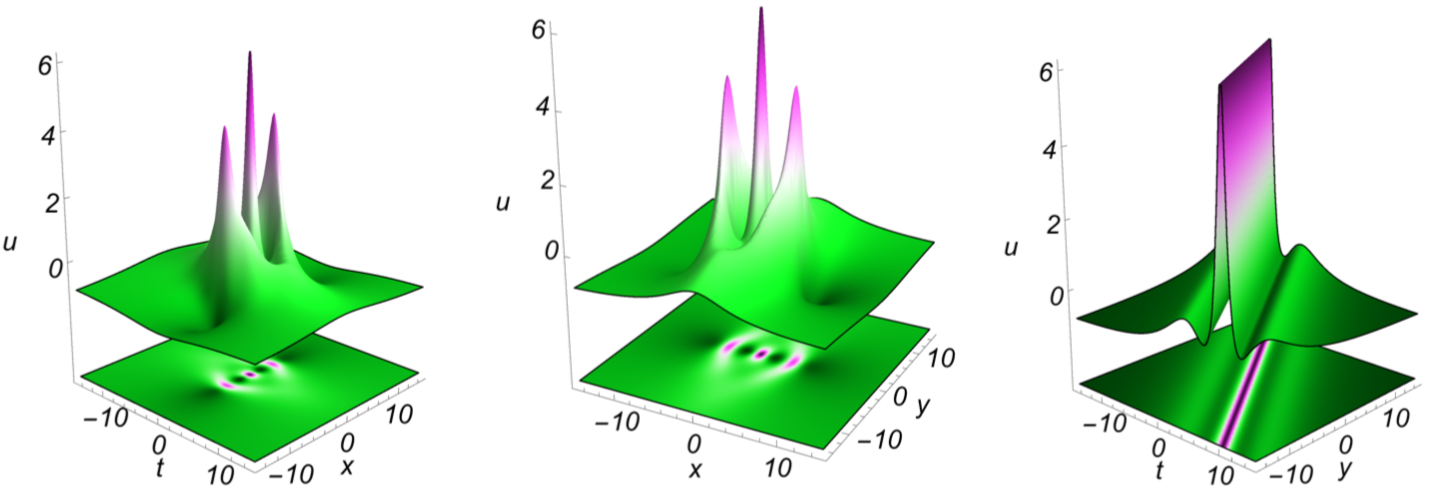}
		\caption{The third order bright rogue waves along $x-t$ and $x-y$ planes at $y=0.05$ and $t=0.05$ with multi-peak doubly localized structures. A multi-peak rational soliton in $y-t$ plane at $x=0.05$. Other parameters are chosen as $u_0 = -0.9, \gamma = 1.5, \beta = 1.5,  \lambda= 0.2, a_1 = 1.5, \mu = 0.05$ and $\alpha = 0.5$.} 
	\end{center} \label{fig-rogue3-1}
\end{figure}
\begin{figure}[h] 
	\begin{center} 
		\includegraphics[width=1.05\linewidth]{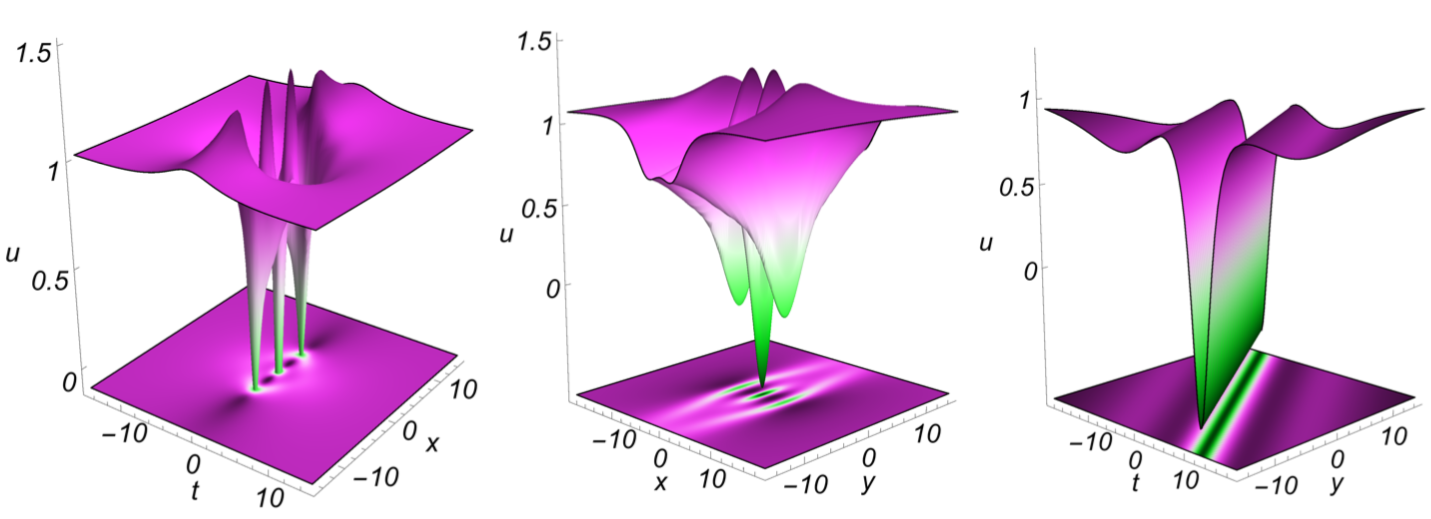}
		\caption{ The third order dark rogue waves along $x-t$ and $x-y$ planes at $y=0.05$ and $t=0.05$ with multi-hole doubly localized structures. A multi-hole with central deep hole rational soliton in $y-t$ plane at $x=0.05$. Other parameters are chosen as $u_0 = 1.04, \gamma = 0.15, \beta = -0.55, \lambda= 0.2, a_1 = 1.2, \mu = 0.5$, and $\alpha = -0.5$.} 
	\end{center} \label{fig-rogue3-2}
\end{figure}

It is a straightforward exercise to construct an arbitrary $N$-th order rogue wave solutions ($N\geq 4$) by following the similar procedure given above. Still, as it involves a tedious task and with lengthy mathematical relations, which restricted us to present/discuss them here in this manuscript.  

\section*{Remarks}
\begin{itemize}
	\item It should be mentioned that the occurrence of $N$-soliton solution ($N\geq 3$) is an alternative approach to confirm the integrability nature of a model. But here, the occurrence of $N$-rogue wave solutions ($N\geq 3$) does not guarantee the integrability \cite{zzh}. Precisely, the $N$-rogue wave solutions ($N\geq 3$) for any model (integrable/non-integrable) can be obtained if it admits the following type of bilinear form \cite{zzh,wli,zzh2}:
	\begin{eqnarray} \label{eq14b}
	(A_1 {D_x^4} +A_2 {D_\tau^2}  +A_3  D_x^2 ) \mathcal{R} \cdot \mathcal{R} =0 ,
	\end{eqnarray}
	where $A_1, A_2$, and $A_3$ are parameters associated with the considered system,  and if the bilinear from is free from any mixed partial derivatives, as mentioned in \cite{zzh2}. In this direction, it would be an interesting study to find out, which kind of bilinear form can show multiple rogue wave solution.
	
	\item The present generalized computation approach is better than the method proposed in \cite{ylm}, because of the fact that the current solutions utilizes mathematical expression to generate the polynomial test function for constructing any arbitrary $N$-th order rogue wave solutions. But, in the case of Ref. \cite{ylm}, two different test functions are necessary to construct rogue wave solutions of order one and two only. 
	
	\item The rogue wave parameters are calculated by the obtained determined/overdetermined systems. In the case of first order rogue wave, we have obtained the determined system while in construction of second and third order rogue wave we have obtained the overdetermined systems. On the other hand, the number of arbitrary parameters affecting the evolution mechanism of rogue waves are seven, nine and seven for first, second and third order rogue wave solutions respectively.
	
	\item  It is interesting to point out that all the obtained rogue wave solutions are having the characteristics $\lim_{|x|\rightarrow \infty} u(x, y, t) = u_0$ and $\lim_{|y|\rightarrow \infty} u(x, y, t) = u_0$.
	
\end{itemize}

\section{Conclusions}\label{sec4} 
We have considered a new integrable (2+1)-dimensional Boussinesq equation proposed recently and constructed higher-order rogue wave solutions by utilizing a generalized polynomial test function implemented through Bell polynomial and Hirota's bilinearization. The dynamics of rogue waves are investigated by carrying out a categorical analysis of the obtained solutions and found the existence of two types of bright and dark type rogue waves. We have explored the evolution dynamics of these rogue waves along with W-shaped, M-shaped, and multi-peak rational solitons. We found that the arbitrary parameters ($u_0$, $\beta$, $\gamma$, $\lambda$, $\alpha$, $\mu$, and $a_1$) available in the obtained solutions help to manipulate the dynamics of rogue waves which enable one to control the amplitude/depth, width, tail-depth, and localization of the bright and dark rogue waves. Also, we found that $a_1$ parameter enacts the transformation of the rogue wave into stationary rational solitons in first as well as all the higher-order solutions. Further, the position shifting of the localized structures along a particular dimension/direction is also possible through these parameters. The results presented in this work will be helpful to the studies on the rogue waves on other higher dimensional systems. Also, it will be helpful to various experimental investigations on the controlling mechanism of rogue waves in optical systems, atomic condensates, deep water oceanic waves, and other related coherent wave systems.

\section*{Acknowledgements}
Sudhir Singh would like to thank Ministry of Human Resource Development (MHRD), Government of India, and National Institute of Technology, Tiruchirappalli, India, for financial support through institute fellowship. KS was supported by Department of Science and Technology - Science and Engineering Research Board (DST-SERB), Government of India, sponsored National Post-Doctoral Fellowship (File No. PDF/2016/000547).\\

\noindent{\bf Declaration}\\
\noindent{\bf Conflict of Interest}: The authors declare that there is no conflict of interests regarding the research effort and the publication of this paper.\\~\\
\noindent{\bf CRediT Author Statement:} Sudhir Singh: Conceptualization, Methodology, Writing - original draft preparation, Writing - review and editing. Lakhveer Kaur: Methodology, Writing - original draft preparation. K. Sakkaravarthi: Formal analysis and investigation, Visualization, Writing - original draft preparation, Writing - review and editing. K.Murugesan; Resources, Funding acquisition, Supervision, Project Management. R.Sakthivel: Resources, Supervision, Project Management.

\section*{Appendix A}\label{secapp}
{Here we briefly revisit the development of the generalized solution structure, which we have adopted for constructing higher-order rogue waves. The rogue wave solution of the simple Boussinesq equation (\ref{eq12}) arising for the choice $\alpha=0$, $\beta=3$, and $\gamma=1$ is reported in Ref. \cite{mja} as
	\bea
	u(x,t) & = 2  \frac{\partial ^2}{\partial x ^2} ln (x^2 - t^2 - 3 ) \Rightarrow   \frac{-4(3+x^2 + t^2)}{(3-x^2 + t^2 )^2}. \label{eq21}
	\eea
	Further, the following form of rational solution for the classical Boussinesq model is proposed in \cite{pac} :
	\bes\bea
	u_r (x,t) = 2  \frac{\partial ^2 }{\partial x ^2 } ln R_ r (x,t) , \qquad r \geq 1, \label{eq22}
	\eea
	where $R_r (x,t)$ is a polynomial in $x^2 $ and $t^2 $ of degree $r(r+1)/2$ and it is expressed as:
	\bea
	R_r (x,t) = \sum \limits _{n=0} ^{r(r+1)/2} \sum \limits _{i=0} ^{n} c_{i,n} x^{2i} t^{2(n-i)}, \label{eq23}
	\eea\ees 
	where $c_{i,n} $ are the constant to be determined by solving equations arising as the coefficients of different powers of $x$ and $t$. Further, the generalization of rational solution is achieved in Ref. \cite{pac} and its explicit form can be written as
	\bes
	\bea 
	\tilde{u}_r (x,t;\lambda , \mu ) = 2  \frac{\partial ^2 }{\partial x ^2 } ln \, \tilde{R}_r (x,t;\lambda, \mu ), \qquad \mbox{for} \quad r \geq 1, \label{eq24}
	\eea
	where
	\bea
	\hspace{-01.5cm}\tilde{R} _{r+1} (x,t;\lambda, \mu ) &=& R _{r+1} (x,t) + 2 \lambda t F_r (x,t) + 2 \mu x G_r (x,t) + (\lambda^2 + \mu ^2 ) R _{r-1} (x,t). \label{eq25}
	\eea
	Here $R_r(x,t)$ take the form of as given in (\ref{eq23}), while $F_r (x,t)$ and $G_r (x,t)$ is obtained as 
	\bea
	\hspace{-01.5cm}F_r (x,t)  = \sum \limits _{n=0} ^{r(r+1)/2} \sum \limits _{i=0} ^{n} e _{i,n} x ^{2i} t^{2(n-i)}, \quad
	G_r(x,t)  =  \sum \limits _{n=0} ^{r(r+1)/2} \sum \limits _{i=0} ^{n} h_{i,n} x^{2i} t^{2(n-i)}.
	\label{eq26}
	\eea \ees
	
	Moving forward, based on the Refs. \cite{mja,pac}, much generalized solution structure with more number of control parameters is proposed recently to construct the higher-order rogue waves for Kadomtsev-Petviashvili (KP) type equation \cite{azh} as
	\bea 
	\mathcal{R} = R_{r+1} (x,\tau) + 2 \lambda \tau F_r (x,\tau) + 2 \mu x G_r (x,\tau) + ( \lambda^2 +\mu ^2 ) R _{r-1} (x,\tau). \label{eq27aa}
	\eea
}

{\setstretch{1.150}
	}}
\end{document}